\documentclass[page-classic]{epl2} 
\usepackage{graphicx}
\usepackage{float}
\usepackage{color}
\usepackage{booktabs}
\usepackage{multirow}
\usepackage{latexsym}
\usepackage{setspace}
\title{Correlation of Crystal Quality and Extreme Magnetoresistance of WTe$_2$}

\author{Mazhar N. Ali\inst{1} \and Leslie Schoop\inst{1} \and Jun Xiong\inst{2} \and Steven Flynn\inst{1} \and Quinn Gibson\inst{1} \and Max Hirschberger\inst{2} \and N. P. Ong\inst{2} \and R. J. Cava \inst{1}}
\shortauthor{M.N. Ali \etal}

\institute{                    
  \inst{1} Department of Chemistry, Princeton University, Princeton New Jersey 08544, USA.\\
  \inst{2} Joseph Henry Laboratories and Department of Physics, Princeton University, Princeton, NJ 08544, USA.
}


\abstract{
\indent{}High quality single crystals of WTe$_2$ were grown using a Te flux followed by a cleaning step involving self-vapor transport. The method is reproducible and yields consistently higher quality single crystals than are typically obtained via halide assisted vapor transport methods. Magnetoresistance (MR)values at 9 Tesla and 2 Kelvin as high as 1.75 million \%, nearly an order of magnitude higher than previously reported for this material, were obtained on crystals with residual resistivity ratio (RRR) of approximately 1250. The MR follows a near B$^2$ law (B = 1.95(1)) and, assuming a semiclassical model, the average carrier mobility for the highest quality crystal was found to be ~167,000 cm$^2$/Vs at 2 K. A correlation of RRR, MR ratio and average carrier mobility ($\mu_{avg}$) is found with the cooling rate during the flux growth.  
}

\begin{document}

\maketitle

\section{Introduction}
\indent{}Recently the layered transition metal dichalcogenide WTe$_2$ was reported to have an extremely large, non saturating magnetoresistance (XMR) \cite{ali2014large} reaching values of 13 million \% at 0.53 Kelvin (K) and 60 Tesla (T). Larger than the giant MR (GMR) or colossal MR (CMR) effects previously seen \cite{GMRpap, CMRcava, CMRpap}, this makes WTe$_2$ of interest due to both its fundamental materials physics and its potential for low temperature applications in magnetic sensors or computing-related devices. Following this work, additional studies on WTe$_2$ have been conducted with findings including pressure-induced superconductivity \cite{pan2015pressure, kang2015superconductivity}, orbital spin texturing \cite{jiang2015signature}, a large and linear Nernst effect \cite{zhu2015quantum}, predictions of a quantum spin hall insulating state under strain \cite{qian2014quantum}, and anomalous phonon behavior with temperature \cite{kong2015raman}. Several other materials such as Cd$_3$As$_2$ \cite{liang2014ultrahigh}, NbSb$_2$ \cite{wang2014anisotropic}, TaAs \cite{zhang2015tantalum}, NbAs \cite{ghimire2015magnetotransport}, and NbP \cite{shekhar2015extremely}, have also recently been reported to show a similarly large MR effect. Ranging in fundamental nature from perfectly compensated semi-metals (WTe$_2$) to Dirac (Cd$_3$As$_2$) to Weyl (TaAs, NbAs, NbP) semi-metals, the origin of this behavior remains under discussion.

\indent{}Experimental studies hoping to probe intrinsic XMR effects are highly sensitive to crystal quality, as can be inferred from the case of WTe$_2$ where there have already been several reports with conflicting data and interpretations. In recent studies, for example, as few as 4 and as many as 9 electron and hole pockets have been identified in complicated ``russian doll''-like Fermi surfaces with pockets nested inside of pockets. \cite{zhu2015quantum, xiang2015multiple, cai2014drastic, PhysRevLett.113.216601, jiang2015signature} While all studies agree on the fundamental importance of near-perfect compensation as the origin of the XMR for WTe$_2$, the need for consistent and reproducible results on high quality crystals is clear. This necessity has not yet been met; Table \ref{table1} shows the XMR values (and, when available, the Residual Resistivity Ratios) of crystals used in various recent investigations of the electronic properties of WTe$_2$. Using the metric of the value of the XMR at ~2 K and 9 T, it can be inferred that there is a large variation in sample quality, leading to quite different conclusions concerning the Fermi surface. 

\indent{}In this work we address the issue of the correlation of the magnitude of the titanic effect to the crystal quality in WTe$_2$. In order to contrast different materials with WTe$_2$, and to compare different batches of WTe$_2$, we believe that it is most useful to compare XMR values at 2 K and 9 T, as these conditions are widely available and have been cited in the various publications. WTe$_2$, first reported to have an MR of 173,000 \% at 2 K and 9 T, was thought to have been surpassed by subsequently characterized materials such as TaAs and NbP, which were reported to have MR's of 540,000 \% and 850,000 \% at ~2 K and 9 T. Here we show, however, that the XMR of WTe$_2$ in higher quality crystals can be an order of magnitude higher than previously reported, specifically in crystals grown via a flux method with a vapor transport cleaning step. WTe$_2$ holds the current record if grown this way (an MR of 1,750,000 \% at 2 K and 9 T was achieved), and, when grown by the flux method described, the results are also reproducible from crystal to crystal within a growth batch. We show that the crystal quality affects the MR strongly; the titanic effect is shown to increase non-linearly with increasing crystal quality as measured by the residual resistivity ratio.  

%Table 1%%%%%%%%%%%%%%%%%%%%%%%%%%%%%%%%%%%%%%%%%%%%%%%%%%%%%%%%%%%%%%%%%%%%%%%%%%%%%%%%%%%%%%%%%%%%%%%%%%%%%%%%%%%%%%%
\begin{table}
\caption{Variation of the titanic MR effect in different studies on WTe$_2$.}
\label{table1}
\begin{center}
\scalebox{0.65}{
\begin{tabular}{lcccr}
\hline
\hline
RRR & MR (ratio) & Applied field (T) & Notes & Citation  \\
\hline
370  & 1730 & 9    & Initial report of Titanic MR in WTe$_2$               & Ali et. al. \cite{ali2014large}  \\
24.6 & 13.4 & 9    & 4 Pockets; 2 electron and 2 hole pockets              & Pletikosic et. al. \cite{PhysRevLett.113.216601}  \\
N/A  & 4538 & 12   & Anomalous phonon mode dependance with temperature     & Kong et. al. \cite{kong2015raman}  \\
50   & 48   & 9    & Superconductivity seen around 11 GPa                  & Pan et. al. \cite{pan2015pressure}  \\
1256 & 31000 & 10  & 8 Pockets; 4 electron and 4 hole pockets              & Zhu et. al. \cite{zhu2015quantum}  \\
N/A  & 18.5 & 8    & 6 Pockets; 2 electron and 4 hole pockets with perfect compensation & Xiang et. al. \cite{xiang2015multiple}  \\
184  & 1250 & 14.5 & 8 Pockets; 4 electron and 4 hole pockets              & Cai et. al. \cite{cai2014drastic} \\
58   & 50   & 9    & 9 Pockets; 4 electron and 5 hole pockets              & Jiang et. al. \cite{jiang2015signature} \\
N/A  & 50   & 7    & Superconductivity seen around 10.5 GPa                & Kang et. al. \cite{kang2015superconductivity} \\
741  & 11322 & 14.7 & Flux growth and Linear MR when B parallel to \textit{a}-axis & Zhao et. al. \cite{zhao2015anisotropic} \\
900  & 6500 & 9    & 6 Pockets; 2 electron and 4 hole pockets, flux growth & Wu et. al. \cite{wu2015temperature} \\
50  & 48   & 9    & 9 Pockets; 4 electron and 5 hole pockets, linear MR & Pan et. al. \cite{pan2015robust} \\
\end{tabular}}
\end{center}
\end{table}

\section{Experimental}
\indent{}The high quality crystals of WTe$_2$ were grown out of a Te flux taking advantage of the fact that W (and correspondingly) WTe$_2$ are very slightly soluble in Te. 200 mg of WTe$_2$ powder (made in a first step as previously reported \cite{ali2014large} and 10 g of purified Te (99.9999\%) were sealed in a quartz tube. This method also works if simply using W powder as a starting material, however an increase in consistency of crystal quality was observed when using a pre-reacted WTe$_2$ precursor. A small amount of quartz wool was added in the tube to act as a filter in order to separate the flux material from the crystals in a later step. The tube was then heated to 825$^\circ$C, held there for 24 h and then cooled at a rate of 2-3$^\circ$C/h to 525$^\circ$C. At this temperature the flux was separated from the crystals by inverting the tube and centrifuging. The WTe$_2$ crystals were then put in another vacuum sealed quartz tube and heated to 415$^\circ$C in a tube furnace, with WTe$_2$ crystals on the hot end and the cold end held at 200 C. This was done for 2 days in order to separate excess Te from the crystal surfaces via self-vapor-transport and to anneal the crystals. This step is crucial to consistently obtaining samples of high quality displaying the XMR effect. 

\indent{}The cooling rate was found to be a significant influence on the crystal quality. The data presented here were reproduced in several different crystal batches for each cooling rate, following the same synthesis procedure. From each batch 3 different crystals were measured, with similar XMR values found every time. The crystals' structure was confirmed as  being that of orthorhombic WTe$_2$ by single crystal diffraction on a Bruker Kappa APEX II using graphite-monochromated Mo K$\alpha$ radiation ($\lambda$=0.71073 \AA) at 100~K. Energy Dispersive X-ray Spectroscopy (EDS) analysis and Scanning Electron Microscopy (SEM) images were taken on a FEI Quanta 200 FEG Environmental SEM system. Transport properties were measured on a Physical Property Measurement System (PPMS) from Quantum Design.

\section{Results and Discussion}
\begin{figure}
\scalebox{0.42}{
\onefigure{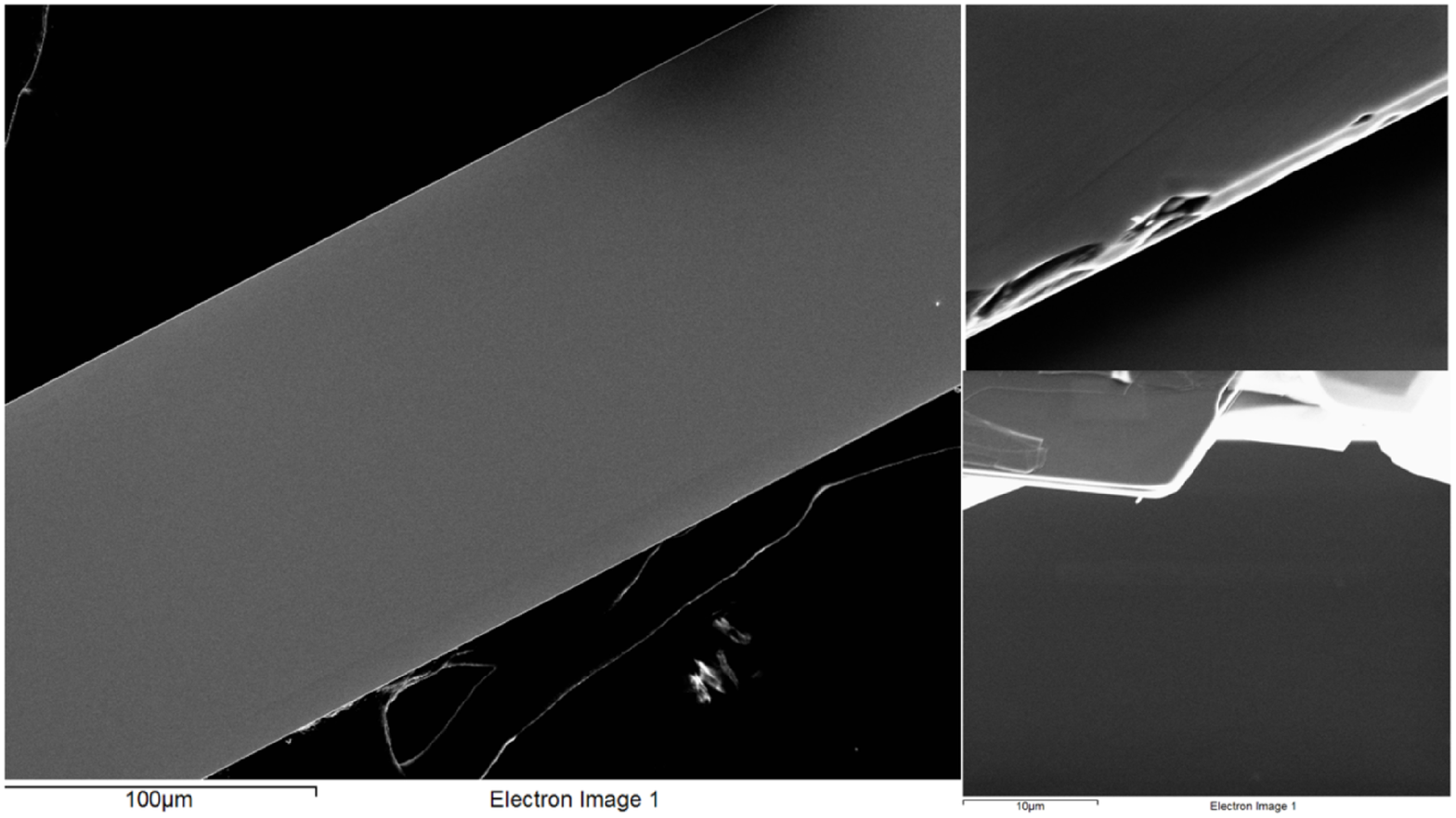}}
\caption{SEM Pictures of WTe$_2$ crystals show very uniform, clean, flat faces (001 direction) with layered growth visible in insets. EDS analysis confirmed the precise stoichiometry of the flux grown and vapor cleaned crystals.}
\label{SEM}
\end{figure}

\begin{figure}
\scalebox{0.57}{
\onefigure{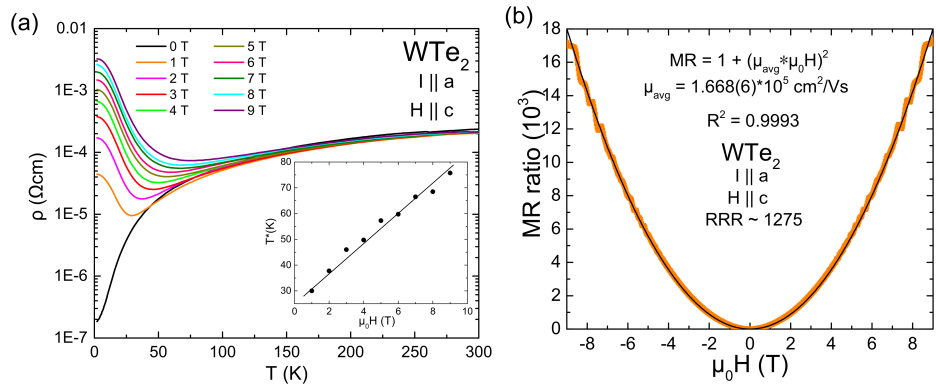}}
\caption{(a) The magnetoresistance of WTe$_2$ grown via the flux method with the current (I) along the \textit{a}-axis and the magnetic field applied (H) along the \textit{c}-axis. The inset shows the dependence of the ``turn on'' temperature (T*) as a function of H. (b) MR vs H at 2 K and the fit to ($\mu$B)$^2$ law. Extrapolated to 60 T, for comparison to \cite{ali2014large}, the MR of these flux grown crystals would approach 100 million $\%$.}
\label{MRdata}
\end{figure}

\indent{}An SEM image of a flux grown crystal is shown in Figure \ref{SEM}; a large clean surface in the \textit{ab} plane, as well as layering along the c axis are clearly visible. The chemical composition was confirmed with EDS to be stoichiometric WTe$_2$. Figure \ref{MRdata}(a) shows the temperature dependent resistivity of a crystal from the same growth batch in different fields. The RRR of these flux grown crystals in zero field reaches a value of 1250; much higher than the values reported for most vapor transport grown crystals (see table \ref{table1}). In general this high RRR value is a good indicator of excellent crystal quality. The resistivity ($\rho$) of the flux grown crystal at 0 T and 2 K is 0.185 $\mu\Omega$-cm; an order of magnitude lower than the resistivity previously reported for WeT$_2$ crystals grown with vapor transport. By 9 T, the resistivity has increased by more than 4 orders of magnitude reaching a peak value of 0.00325 $\Omega$-cm; surprisingly close to the resistivity at 2 K and 9 T of the lower quality crystal previously reported (0.0031 $\Omega$-cm) \cite{ali2014large}. This decrease in zero field resistivity results in the flux grown crystal achieving an MR of 1.75 million percent by 9 T. The inset to Figure \ref{MRdata}(a) shows the linear dependence of T*, the temperature of the minimum in $\rho$ vs. T, as a function of applied field.   

\indent{}By fitting the field dependent resistivity at 2 K (Figure \ref{MRdata}(b)) to the Lorentz law (MR ratio = 1 + ($\mu_{avg}B)^2$ we can extract the average carrier mobility ($\mu_{avg}$), finding a value of 1.67$ \cdot 10^5$ cm$^2/Vs$ for the high quality crystal whose properties are shown in Figure \ref{MRdata}(a). This value is not as high as the ultrahigh carrier mobilities observed in some Dirac and Weyl semimetals such as Cd$_3$As$_2$, TaAs, NbAs and NbP\cite {liang2014ultrahigh, zhang2015tantalum, ghimire2015magnetotransport, shekhar2015extremely}, but it is unusually high for a material that is not believed to have Dirac electrons. The high mobility may arise from two reasons; first the high RRR for these crystals implies the presence of a very small number of defects, therefore reducing the probability of defect/impurity scattering, and second, the band structure of WTe$_2$ shows steep bands around the Fermi level \cite{ali2014large}, which also implies high mobility due to low effective mass charge carriers.

\begin{figure}
\scalebox{0.47}{
\onefigure{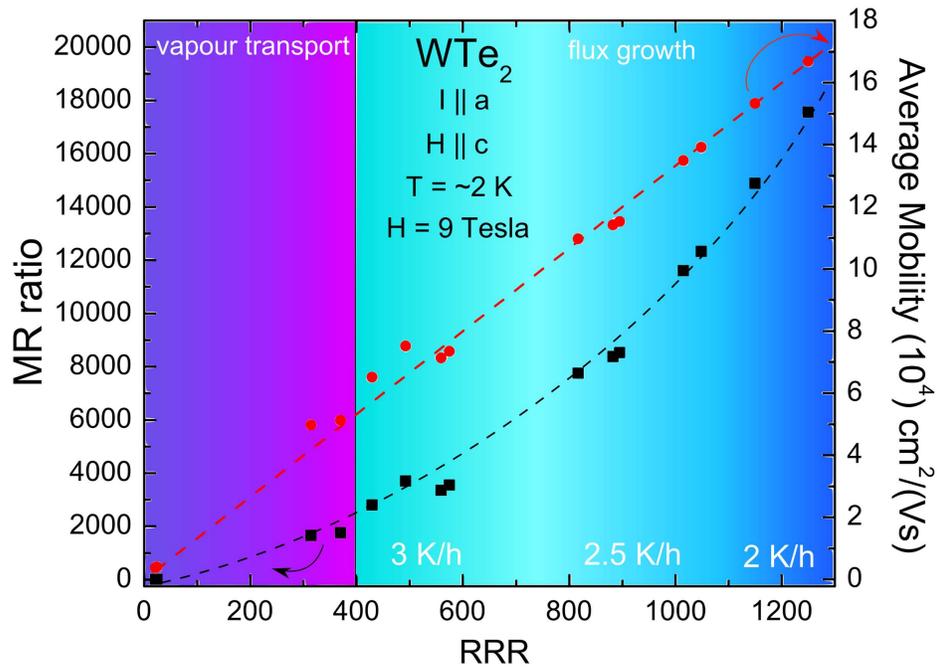}}
\caption{MR ratio and Average carrier mobility versus RRR. MR ratio follows the left axis (black squares) while $\mu_{avg}$ follows the right axis (red circles). The dashed lines are guides to the eye with the black line meant to show the non-linear correlation of the MR ratio with RRR and the red line meant to show the linear correlation of $\mu_{avg}$ with RRR. Shading is used to distinguish crystals made using the flux method at particular cooling rates (3 $^\circ$C/hr, 2.5 $^\circ$C/hr and 2 $^\circ$C/hr) from crystals made using vapor transport.}
\label{RRR}
\end{figure}

\indent{}Finally Figure \ref{RRR} shows the correlation of the MR ratio at 9 T and 2 K and the average carrier mobility (as extracted from Lorentz fitting) with the RRR value for various crystals fabricated by using vapor transport as well as the flux method. This is the primary result of the current study. For the flux method, 3 different cooling rates (given in $^\circ$C/hr) were used to control the RRR value and, correspondingly, the MR and mobility. The color coding indicates the growth method (flux or vapor transport) as well as different cooling rates for different flux growth batches. The highest RRR was obtained with the slowest cooling rate. The correlation of $\mu_{avg}$ with RRR appears roughly linear while the MR appears to follow a roughly square law correlation, consistent with the Lorentz magnetoresistance law. Figure \ref{RRR} also shows how different batches can produce crystals of similar quality, although as the cooling rate slows the consistency between batches decreases. This is likely due to experimental considerations, such as consistency in the programmed cooling rate of the furnace and the centrifugation step, causing small changes in RRR that can drastically affect the MR ratio due to the non-linear dependence. 

\indent{}This not only points out the importance of improving the crystal growth method, but also indicates that the high MR is at least partially the result of relatively defect free crystals. The fact that we were able to grow WTe$_2$ with a RRR of 1250 by a simple lab technique implies that this compound has a very low tendencey to form defects in Te-rich growth conditions. This may be aid in thin film synthesis or larger scale fabrication methods, such as liquid phase epitaxy, for device applications.

\section{Conclusion}
\indent{}
In conclusion, we have described a new crystal growth method for obtaining high quality crystals of WTe$_2$. The method is relatively straightforward to perform and is reproducible. By using a secondary cleaning/annealing step, we were able to consistently control the crystal quality, which is more consistent over the whole batch, compared to vapor transport growth which has larger sample to sample variation within the same batch. Magnetoresistances as high as 1.75 million \% at 9 Tesla and 2 Kelvin were achieved on crystals with RRR values of $~$1250; nearly an order of magnitude higher than previously reported. The MR follows a near B$^2$ law (B = 1.95(1)) and the average carrier mobility was found to be ~167,000 cm$^2$/Vs at 2 K for the high quality crystals. A correlation of RRR, MR ratio and average carrier mobility is found with cooling rate during the flux growth; the slower the crystals are cooled, the higher the RRR and MR ratio. We attribute this to the minimal defects and impurities in the flux grown crystals. Adoption of the technique described here may lead to crystals of higher and consistent quality that will in turn lead to more meaningful comparisons across studies from different research groups as investigations of the electronic properties of WTe$_2$ continue.

\acknowledgments
This research was supported by the National Science Foundation MRSEC program grant DMR-1420541. \newline

\newpage
%%%%%%%%%%%%%%%%%%%%%%%%%%%%%%%%%%%%%%%%%%%%%%%%%%%%%%%%%%%%%%%%%%%%%%%%%%%%%%%%%%%%%%%%%%%%%%%%%%%%%%%%%%%%%%%%%%%%%%%%%%%%%%%%%%%%%%%%%%%%%%%%%%%%%%%%%%
\newpage
%%%%%%%%%%%%%%%%%%%%%%%%%%%%%%%%%%%%%%%%%%%%%%%%%%%%%%%%%%%%%%%%%%%%%%%%%%%%%%%%%%%%%%%%%%%%%%%%%%%%%%%%%%%%%%%%%%%%%%%%%%%%%%%%%%%%%%%%%%%%%%%%%%%%%%%%%%
\newpage

\end{document}